\documentclass[twocolumn,superscriptaddress,amssymb,amsmath,aps,prl]{revtex4-1}
\usepackage{color}
\usepackage{graphicx}
\usepackage{esint}
\usepackage{mathptmx}
\usepackage{balance}

\makeatletter

\@ifundefined{textcolor}{}
{%
 \definecolor{BLACK}{gray}{0}
 \definecolor{WHITE}{gray}{1}
 \definecolor{RED}{rgb}{1,0,0}
 \definecolor{GREEN}{rgb}{0,1,0}
 \definecolor{BLUE}{rgb}{0,0,1}
 \definecolor{CYAN}{cmyk}{1,0,0,0}
 \definecolor{MAGENTA}{cmyk}{0,1,0,0}
 \definecolor{YELLOW}{cmyk}{0,0,1,0}
}

\@ifundefined{textcolor}{}{%
 \definecolor{BLACK}{gray}{0}
 \definecolor{WHITE}{gray}{1}
 \definecolor{RED}{rgb}{1,0,0}
 \definecolor{GREEN}{rgb}{0,1,0}
 \definecolor{BLUE}{rgb}{0,0,1}
 \definecolor{CYAN}{cmyk}{1,0,0,0}
 \definecolor{MAGENTA}{cmyk}{0,1,0,0}
 \definecolor{YELLOW}{cmyk}{0,0,1,0}
}

\makeatother

\begin{document}

	\title{Experimental observation of exceptional surface in synthetic dimensions with magnon polaritons}

	\author{Xufeng Zhang}%
	\email{xufeng.zhang@anl.gov}
	\affiliation{ 
		Center for Nanoscale Materials, Argonne National Laboratory, Argonne, IL 60439, USA
	}%

	\author{Kun Ding}%
	\affiliation{ 
		The Blackett Laboratory, Department of Physics, Imperial College London, London SW7 2AZ, United Kingdom
	}%
	
	\author{Dafei Jin}%
	\affiliation{ 
		Center for Nanoscale Materials, Argonne National Laboratory, Argonne, IL 60439, USA
	}%
	
	\author{Xianjing Zhou}%
	\affiliation{ 
		Center for Nanoscale Materials, Argonne National Laboratory, Argonne, IL 60439, USA
	}%

	\author{Jing Xu}%
	\affiliation{ 
		Center for Nanoscale Materials, Argonne National Laboratory, Argonne, IL 60439, USA
	}%

\date{\today}
\begin{abstract}
	Exceptional points (EPs) are singularities of energy levels in non-Hermitian systems. In this Letter, we demonstrate the surface of EPs on a magnon polariton platform composed of coupled magnons and microwave photons. Our experiments show that EPs form a three-dimensional exceptional surface (ES) when the system is tuned in a four-dimensional synthetic space. We demonstrated that there exists an exceptional saddle point (ESP) in the ES which originates from the unique couplings between magnons and microwave photons. Such an ESP exhibits unique anisotropic behaviors in both the real and imaginary part of the eigenfrequencies. To the best of our knowledge, this is the first experimental observation of ES, opening up new opportunities for high-dimensional control of non-Hermitian systems.
		
\end{abstract}

\maketitle


Exceptional point (EP) is the singularity of coupled dissipative systems which can be described by a non-Hermitian Hamiltonian\cite{2012_JPA_Heiss}. At the EP, both the eigenstates and eigenvalues of the system coalesce, distinguishing it from regular mode degeneracy. EPs have been studied in a wide range of systems, including coupled resonators or waveguides in the optical\cite{2019_Science_Alu,2018_Nature_Yoon,2017_Nture_Hodaei,2017_Nature_Chen}, microwave\cite{2001_PRL_Dembowski,2003_PRL_Dembowski,2004_PRE_Dembowski,2016_Nature_Doppler}, magnetic\cite{2017_PRA_Zhang}, or mechanical domains\cite{2016_Nature_Xu,2016_PRX_Ding}. Novel properties have been discovered around or at the EP, ranging from topological mode transfer\cite{2016_Nature_Xu} and asymmetric mode conversion\cite{2016_Nature_Doppler,2018_Nature_Yoon,2011_JPA_Uzdin} to extraordinary sensitivities\cite{2017_Nture_Hodaei,2017_Nature_Chen} and directional lasing\cite{2016_PNAS_Peng}.

Nonetheless, the experimental demonstrations of EPs have been limited to isolated points or lines of EPs. It has been theoretically proposed recently that surfaces of EPs can be obtained in high dimensional systems\cite{2019_Optica_Zhen,2019_PRB_Okugawa}, which can enable intriguing physical phenomena. However, realization of such exceptional surfaces (ESs) requires more degrees of freedom and great tunabilities, which pose a significant challenge for the experimental demonstration.

On the other hand, magnonic polaritons\cite{2014_PRL_Zhang, 2014_PRL_Tabuchi, 2014_PRApplied_Goryachev, 2015_PRL_Bai, 2013_PRL_Huebl, 2017_SciRep_Bhoi} has been emerging as a promising platform for non-Hermitian physics\cite{2017_NComm_JQYou}. Strong coupling and even ultrastrong coupling between magnons--quantization of collective spin excitations--and microwave photons have been demonstrated. Different from most recently demonstrated non-Hermitian systems, magnon polaritons couple two resonances of different physical natures: spin excitations and electromagnetic waves. In such systems, the magnon frequency can be tuned in a broad range by an external magnetic field, while the coupling strength is determined by the geometry configuration. With such prominent flexibility, magnon polaritons provide an ideal solution for the experimental realization of ES.

In this Letter, we show that, using a magnon polariton system with multiple tuning parameters, EPs can form a three-dimensional (3D) ES within a four-dimensional (4D) synthetic space. Synthetic dimensions have recently been introduced to generate novel topological states that are otherwise difficult to realize\cite{2019_Nature_Lustig,2017_PRX_Wang,2018_Optica_Yuan}. Furthermore, the ES can be conveniently tuned in multiple dimensions simultaneous to coalesce into an exceptional saddle point (ESP). Our measurements show that this ESP is indeed an anisotropic EP, where the real and imaginary part of the eigenfrequencies behave differently along three synthetic dimensions. This is distinctly different from recent demonstrations in low-dimensional parameter spaces\cite{2018_PRL_Ding}, where an EP pair coalesces in a single dimension and forms a EP with anisotropic behavior only in the imaginary part of the eigenfrequencies. The remarkable observation from our study can trigger the study of synthetic dimension EPs in different systems and their applications in topological state transfer and sensing.

\begin{figure}[tb]
	\centering
	\includegraphics[width=1.00\linewidth]{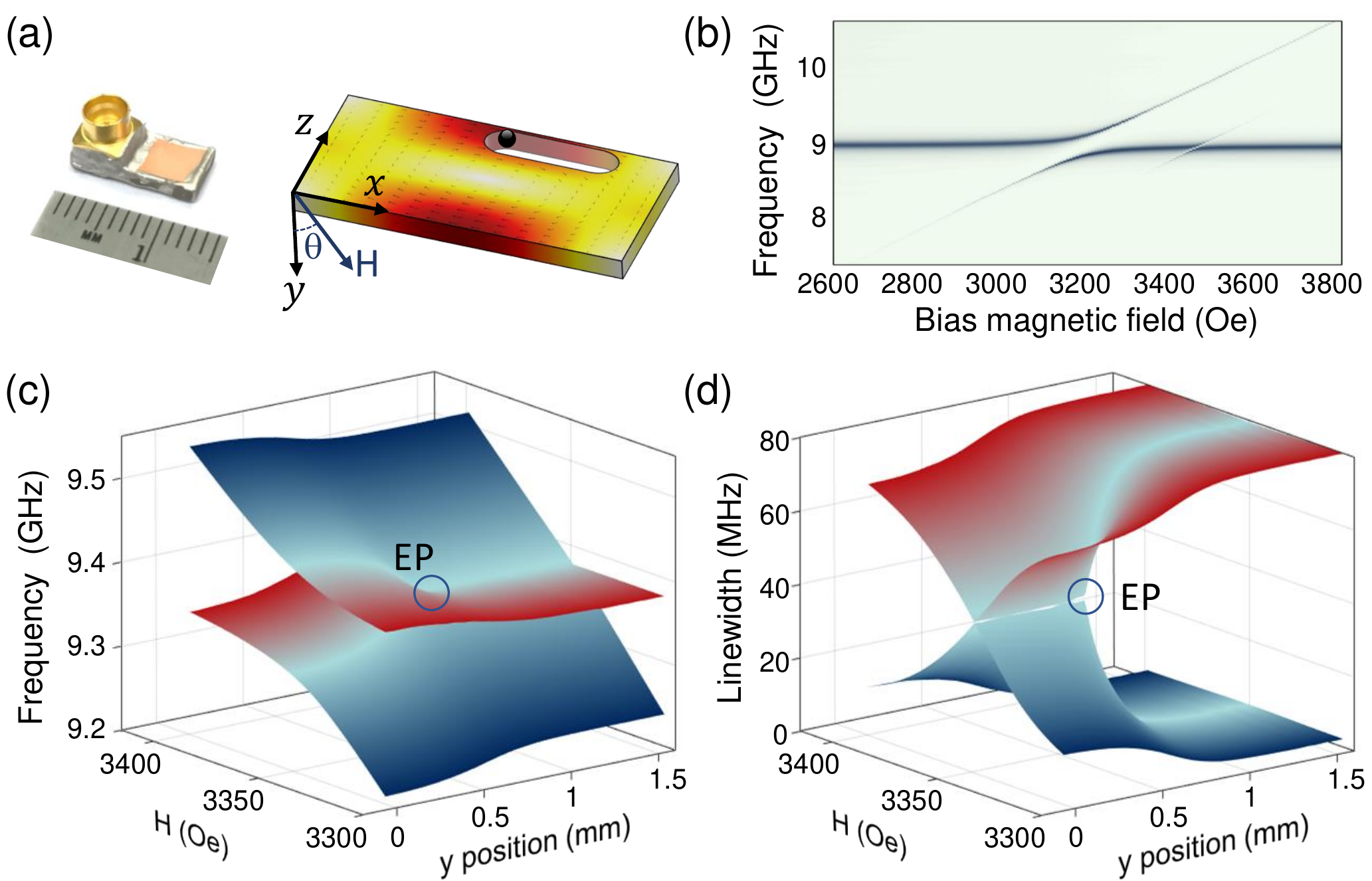}
	\caption{(a) Optical image of the TMM10i microwave cavity and the simulated microwave magnetic field of the cavity TE$_{101}$ mode (fields inside the slot not shown). The circular structure on the left hand side of the cavity is the SMP connector. A slot is drilled near the field maximum to host the YIG sphere. The bias magnetic field is applied in the $x-y$ plane with an angle $\theta$ to the $y$ direction. (b) Normalized reflection spectra from the cavity at different bias fields, showing the anti-crossing. A high order mode is visible on the right hand side of the main anti-crossing. The slot length is 1mm. (c)\&(d) Riemann surfaces for the real (resonance frequency) and imaginary (resonance linewidth) part of the eigenfrequency, respectively, reconstructed from measured reflection spectra and averaged over multiple measurements. EP is indicated by the circle, showing the bifurcation of the Riemann surface.}
	\label{fig:fig1}
\end{figure}

Our system consists of a microwave cavity and a highly polished yttrium iron garnet (YIG, Y$_3$Fe$_5$O$_{12}$) sphere which supports magnon resonances. The microwave cavity is a piece of high-dielectric constant PCB substrate (TMM10i\cite{Rogers}) with doubled-sided copper cladding and enclosed by silver epoxy on the four sidewalls (Fig.\,\ref{fig:fig1}a). One SMP connector is glued on the cavity to probe the microwave resonance, with its metal pin inserted inside a 1-mm diameter hole drilled into the substrate. The cavity is designed to have its TE$_{101}$ mode (Fig.\,\ref{fig:fig1}a) resonating at around 9 GHz, with a loaded linewidth of 94 MHz. The broad linewidth stems from the large dielectric and metal losses as well as the strong perturbation from the SMP probe. However, because of the large dielectric constant of the TMM10i board ($\varepsilon=9.8$), the cavity size is only $12\times 1.2\times 5$ mm$^3$, which is significantly reduced compared with conventional air-filled 3D microwave cavities operating at similar frequencies\cite{2014_PRL_Zhang,2014_PRL_Tabuchi}. Such volume reduction improves the mode matching between the magnon mode and the cavity mode. As a result, the magnon-photon coupling is drastically boosted and strong coupling can be easily achieved.

The YIG sphere we used has a diameter of 400 $\mu$m and is glued on a ceramic rod. A 1-mm-wide slot is cut through the TMM10i board to host the YIG sphere inside the cavity. The slot length varies in different measurement, causing slight changes in the cavity resonance frequency and magnon-photon coupling strength. The YIG sphere is mounted on a two-axis translational stage and can move inside the slot. A bias magnetic field is applied in the $x-y$ plane with a tunable angle ($\theta$) from the $y$ direction, but if not specified, the measurements described in this Letter use a magnetic field parallel to the $y$ direction. The magnon mode of interest, i.e., the ferromagnetic resonance (FMR) mode of the YIG sphere, is tuned by the bias magnetic field $f_m=\gamma H$ where $\gamma=2.8$ MHz/Oe is the gyromagnetic ratio. When the magnon is tuned to near resonance with the cavity mode, they couple with each other through magnetic dipole-dipole interaction. 

\begin{figure}[t]
	\centering
	\includegraphics[width=0.9\linewidth]{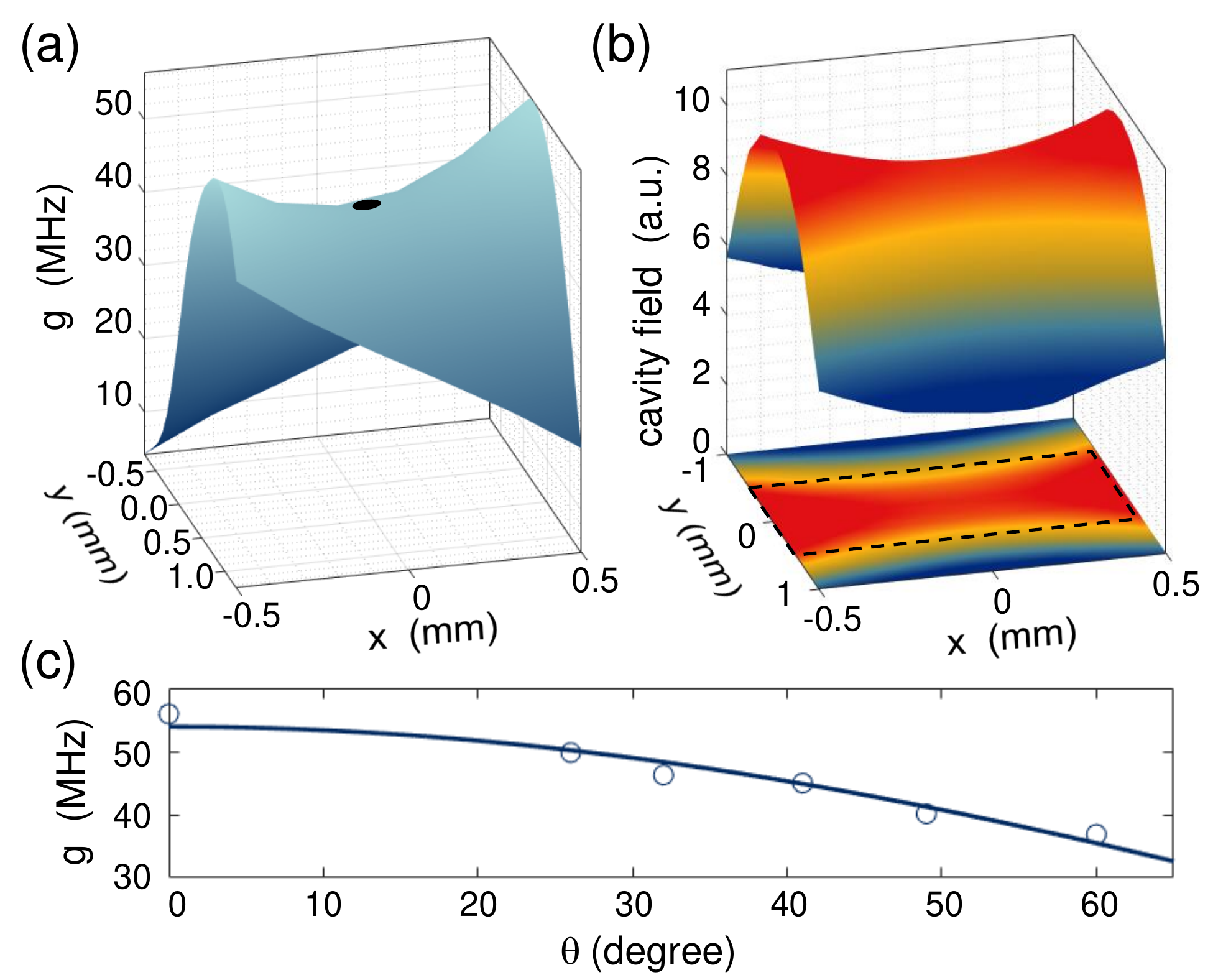}
	\caption{(a) Coupling strength as a function of $x$ and $y$ position of the YIG sphere. The black dot indicates the saddle point. The slot length is 5 mm. (b) Simulated microwave magnetic field distribution for the cavity mode around the slot. Dashed black line indicates the slot area. (c) Coupling strength as a function of $\theta$ at $x=0$ mm and $y=0.05$ mm, where $\theta$ is the angle of the bias magnetic field relative to the $y$ direction. Circles and the solid line are the measurement and cosine fitting results, respectively.}
	\label{fig:fig2}
\end{figure}

The magnon-microwave photon coupling is characterized by measuring the cavity reflection spectrum at different magnetic fields. Avoided-crossing spectra are obtained (Fig.\,\ref{fig:fig1}b) using a cavity with a slot length of 1 mm, clearly showing the strong coupling between the magnon and the cavity photon modes. Intrinsic system parameters are extracted via numerical fittings: cavity resonance frequency $f_c=8.977$ GHz, cavity mode dissipation rate $\kappa_c=54$ MHz, magnon mode dissipation rate $\kappa_m=1.1$ MHz, coupling strength $g=128$ MHz. It is evident that the magnon-microwave photon coupling dominates over the dissipation rates of the magnon and cavity photon modes ($g>\kappa_c,\kappa_m$), confirming the strong coupling between them. A large cooperativity of $C=\frac{g^2}{\kappa_c\times\kappa_m}=276$ is obtained.

\begin{figure*}[tb]
	\centering
	\includegraphics[width=0.8\textwidth]{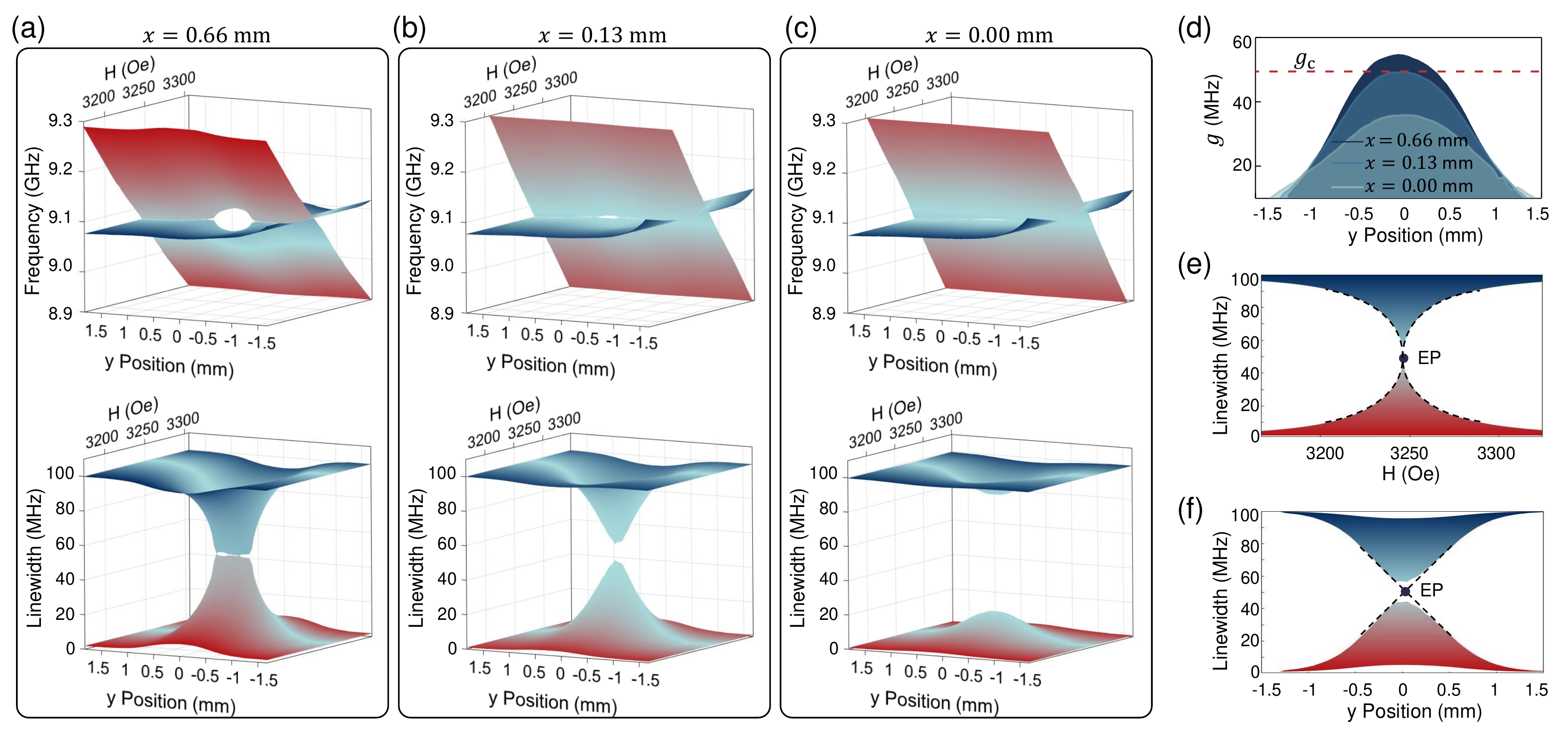}
	\caption{(a)-(c) Experimentally obtained Riemann surfaces for the real (top) and imaginary (bottom) part of the eigenfrequencies at different $x$ positions of the YIG sphere: $x=0.66$ mm ($g>g_c$), $x=0.13$ mm ($g=g_c$), $x=0.00$ mm ($g<g_c$). EP coalescence is observed at $x=0.13$ mm when $g=g_c$. (d) Coupling strength $g$ as a function of $y$ positions at various $x$ locations.  (e) \& (f) Cross-sectional views at the EP for the lower figure of (b). Dashed lines are plotted to guide the eye.}
	\label{fig:fig3}
\end{figure*}

In general, our magnon polariton system can be described by a non-Hermitian Hamiltonian: 
\begin{equation}
H= \left( \begin{matrix} f_c&0\\ 0&f_m \end{matrix}\right)+\left(\begin{matrix} -i\kappa_c&g\\g&-i\kappa_m	\end{matrix} \right).
\label{Eq:Hamiltonian}
\end{equation}
Solving the Hamiltonian gives two eigenmodes at eigenfrequencies:
\begin{equation}
\lambda_\pm=f_\pm+i\kappa_\pm=f_0+i\kappa_0\pm\frac{1}{2}\sqrt{(\Delta f-i\Delta\kappa)^2+4g^2},
\label{Eq:eigenfreq}
\end{equation}
\noindent where $f_0=(f_c+f_m)/2$, $\kappa_0=(\kappa_c+\kappa_m)/2$, $\Delta f=f_m-f_c$, $\Delta\kappa=\kappa_m-\kappa_c$. The parameters needed for calculating the eigenfrequencies can be extracted from numerical fitting of the cavity reflection spectra\cite{SM}.

Equation\,(\ref{Eq:eigenfreq}) indicates that for the two-mode coupled system, there exist two eigenmodes. The Riemann surfaces in Figs.\,\ref{fig:fig1} (c)\&(d) show the real (resonance frequency) and imaginary (linewidth) part of the eigenfrequencies, respectively, as a function of bias magnetic field and $y$ position of the YIG sphere. These Riemann surfaces are obtained from calculations based on experiment results and averaged over multiple measurements to eliminate noises. For each ($H$, $y$) combination, there are two eigenfrequencies. For large $y$ values, the YIG sphere is far away from the slot and therefore the mode overlap between magnon and cavity photon modes is small, resulting in a small coupling strength. As a result, the eigenfrequencies represent the intrinsic magnon and cavity photon modes, with the real part crossing each other at the diabolic points while the imaginary part separated from each other. When $y$ becomes smaller, YIG sphere is closer to the center of the slot along $y$ direction. The increased mode overlap leads to strong coupling, with avoided crossing in the real part of the eigenfrequencies and mixing of the imaginary part (linewidth). However, at the onset of the strong coupling when $y=0.7$ mm and $H=3348$ Oe, there exist only one eigenfrequency. This corresponds to the singularity condition in Eq.\,(\ref{Eq:eigenfreq}): $\Delta f=0$ and $g=g_c=\Delta\kappa/2$, where the two eigenfrequencies coalesce into one. Such a singularity on the Riemann surfaces (for both the real and imaginary part of the eigenfrequencies) is referred to as an EP. It is different from the two-mode degeneracy which has the same eigenfrequecy but different eigenfunctions. Instead, at the EP the two eigenfunctions also coalesce into one. Note here only a half of the Riemann surface is plotted for clarity. Since moving YIG sphere along either direction of $y$ axis is equivalent, the Riemann surface is symmetric along $y$ axis, and therefore there exists another EP at $y=-0.7$ mm, forming an EP pair.

\begin{figure*}[tb]
	\centering
	\includegraphics[width=0.7\linewidth]{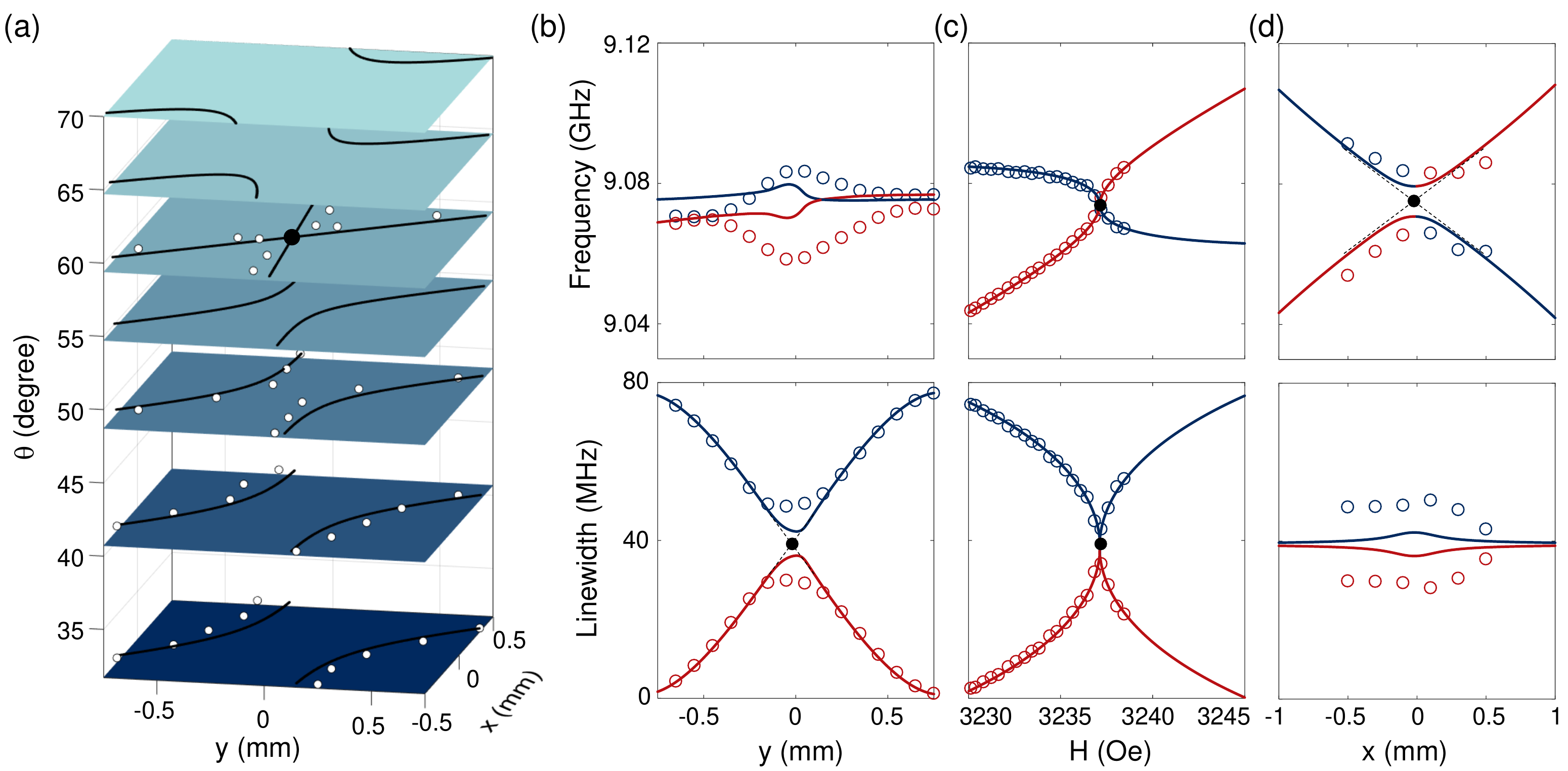}
	\caption{(a) Slices of the ES in the 3D parameter space ($x$,$y$,$\theta$). The fourth dimension $H$ is hidden (fixed at $H=3327$ Oe). White dots are the results directly extracted from experimental data, and black lines are calculated from numerical fittings. The planes at given $\theta$s are guides to the eyes. The black dot represents the ESP. (b)-(d) Cross-sections of the Riemann surfaces at the ESP as a function of $y$ position, magnetic field ($H$), and $x$ position, respectively. Top figures: real part of the eigenfrequencies (resonance frequencies). Bottom figures: imaginary part of the eigenfrequencies (linewidth). Open circles are the experimental results, solid lines are numerical fittings, and dashed lines are to guide the eyes. Black dots represent EPs.}
	\label{fig:fig4}
\end{figure*}


The multi-degrees of freedom in our system allows manipulation of EPs in a high-dimensional synthetic space. In addition to the $y$ position of the YIG sphere and the bias magnetic field $H$, the EP condition is also determined by the $x$ position of the YIG sphere. Figure\,\ref{fig:fig2}(a) plots the measured magnon-microwave photon coupling strength as a function of $x$ and $y$ positions. The decrease in the coupling strength compared with Fig.\,\ref{fig:fig1} is attributed to the extended slot length (5 mm) which gives larger tuning range for $g$ but less confined cavity fields. The saddle-shaped distribution of $g$ is determined by the spatial distribution of the cavity TE$_{101}$ mode inside the slot(Fig.\,\ref{fig:fig2}b). On the other hand, the coupling strength is also affected by the overlapping factor\cite{2014_PRL_Zhang}: $g\propto h\cos\theta$, where $h$ is microwave magnetic field of the cavity mode and $\theta$ is the angle of the applied bias magnetic field relative to the $y$ axis. Note that the tilted bias magnetic field does not affect the magnon frequency because its amplitude does not change. Figure\,\ref{fig:fig2}(c) depicts the experimentally observed coupling strength as a function of $\theta$, showing a clear cosine dependence. Therefore, EPs in our system can be tuned in four synthetic dimensions.


The extra synthetic dimensions introduce convenient control over the magnon polariton EPs. Specifically, an EP pair can coalesce and form an anisotropic EP \cite{2018_PRL_Ding}. Figures\,\ref{fig:fig3}(a)--(c) plots the experimentally obtained Riemann surfaces at different $x$ positions. The slight asymmetry along $y$ position axis is attributed to the fact that the supporting rod of the YIG sphere moves into the slot and perturbs the cavity resonance. The $x$ position of the YIG sphere controls the relative position of the two EPs. When $x=0.66$ mm, the EP pair is clearly visible and widely separated. As $x$ reduces, the two EPs move closer. When $x=0.13$ mm, the two EPs meet and coalesce. As $x$ keeps decrease, no EP can be observed (at $x=0$, e.g.), and the real part of the two eigenfrequencies always crosss each other at diabolic points while the imaginary parts are separated. This can be explained by comparing the coupling strength for different $x$ positions at a bias magnetic field enabling zero detuning $\Delta f=0$ (Fig.\,\ref{fig:fig3}d). When $x$ is large, the maximum of the $g-x$ curve is above the $g=g_c$ line and the two curves cross twice, corresponding to two EPs. Decreasing $x$ also leads to decreases in the couping strength, moving the whole $g-x$ curve below the $g=g_c$ line and therefore they have no intersections, which consequently eliminates any EPs. At $x=0.13$ mm, the maximum of the $g-x$ curve is equal to $g_c$ so the two curves is tangent to each other with a single intersection point at $y=0$ mm, corresponding to a single EP.

A single EP is obtained after the coalescence of the EP pair. For the real part of the eigenfrequency, no signature of mode coupling can observed in the Riemann surface. However, in the imaginary part the singularity condition associated with this EP can be easily observed: the top and bottom surfaces are separated from each other but in contact at a single point. Closer examination at this point shows that this coalesced EP behaves differently along the two parameter axises, as shown in Figs.\,\ref{fig:fig3}(e)\&(f). A linear dependence of the linewidth on the $y$ position is obtained near the EP, while for the bias magnetic field it has a square-root dependence. This can be explained by the fact that the coupling strength $g$ has a parabolic dependence on $y$, while the detuning $\Delta f$ is linearly dependent on $H$.\cite{SM}


However, the EP coalescence only takes place for $y$ positions. When fixing $y=0$ and sweeping $x$, another coalesced EP appears at $x=-0.13$ mm considering the symmetric distribution of the coupling strength $g$ along the $x$ direction (Fig.\,\ref{fig:fig2}a), and these two coalesced EPs form a pair. But within the 4D synthetic space, we can conveniently manipulate the EP pairs to further coalesce them for $x$ positions by taking advantage of the fourth dimension -- the magnetic field angle $\theta$.

Figure\,\ref{fig:fig4}(a) plots the distribution of EPs in the 4D synthetic space ($y$, $x$, $\theta$, $H$). The fourth dimension $H$ is hidden by fixing $H$ at 3237 Oe because EPs always occur at zero detuning ($\Delta f=0$). The EPs calculated from experimental data is represented by the white dots, while theoretical calculations from extrapolated data are shown in solid lines, and a good agreement is observed. These EPs form an ES which exhibits a saddle-shaped distribution and is in agreement with the relation between the coupling strength and the $x$ and $y$ positions.

EPs always form pairs either along $x$ or $y$ axis on such an ES. However, by varying $\theta$, a unique condition can be found when the EP pair coalesces along both axises. Changing $\theta$ effectively changes the overall amplitude of the saddle surface of the coupling strength. When $\theta$ is small (large), the coupling strength is large (small). Therefore, the saddle surface in Fig.\,\ref{fig:fig2}(a) intersects the $g=g_c$ plane below (above) the saddle point, and the intersection is a hyperbola with a gap along $y$ ($x$) direction. Note at these intersections, $g=g_c$ is satisfied, so the parameter combination at these intersections ($x$, $y$, $\theta$, $H$) represents the EP condition in the synthetic space, which is summarized in Fig.\,\ref{fig:fig4}(a). At a critical angle $\theta_c$, the saddle surface intersects the $g=g_c$ plane at the saddle point ($x=x_c$ and $y=y_c$). In this case, the intersection is a single point instead of a hyperbola, indicating the coalescence of EP pairs into one singularity simultaneously in both $x$ and $y$ axis (black dot on the $\theta=60^{\circ}$ plane in Fig.\,\ref{fig:fig4}a). 

The EP coalescence in the 4D synthetic space leads to a nontrivial phenomena: high-dimensional anisotropic EP. Figures\,\ref{fig:fig4}(b)--(d) plot the eigenfrequencies of the system around this ESP. Similar to Fig.\,\ref{fig:fig3}, the EP coalescence results in anisotropic behavior along $y$ and $H$ axis in the imaginary part of the eigenfrequency (linewidth). The linewidth shows a linear dependence on $y$ and a square-root dependence on $H$. While for the $x$ parameter, the anisotropic EP occurs in the real part of the eigenfrequency (resonance frequency). A linear dependence on $x$ is observed for the resonance frequency, which shows a square-root dependence on $H$. Therefore, the ESP is anisotropic along three different synthetic dimensions for both real and imaginary part of the eigenfrequencies. The linear crossings in both real and imaginary parts of the eigenfrequencies at a single EP are unique to our ESP because of the coupling strength distribution shown in Fig.\,\ref{fig:fig2}(a). Such carefully designed coupling conditions offer the degrees of freedom to realize other interesting phenomena in the future, such as anisotropic high-order EPs. From the application point of view, the resonance frequency and linewidth can be used independently to sense different physical variables with various sensitivities and dynamic ranges.

To summarize, we have experimentally demonstrated the ES in a high-dimensional synthetic space for magnonic polaritons. Such an ES can coalesce to an ESP, leading to the emergence of 3D anisotropic behaviors. Our demonstration shows the great potential of magnon polaritons for high-dimensional non-Hermitian physics and opens up new opportunities. For instance, encircling an EP in the high-dimensional synthetic space can enable new functionalities for topological state transfer or unidirectional propagation. In addition, time reversal symmetry breaking in the high-dimensional synthetic space can be obtained considering the magnetic nature of magnons. Moreover, our demonstration can also be extended to magnon-based quantum information processing, where the high-synthetic-dimensional control can enable robust quantum state transduction. Therefore, our results lay the groundwork for magnonic non-Hermitian physics and point out a new avenue for magnon-based signal processing.

This work was performed at the Center for Nanoscale Materials, a U.S. Department of Energy Office of Science User Facility, and supported by the U.S. Department of Energy, Office of Science, under Contract No. DE-AC02-06CH11357. K.D. acknowledges funding from the Gordon and Betty Moore Foundation. X.Z. thanks Dr. Daniel Lopez for enlightening discussions and Dr. Gary Wiederrecht for proofreading the manuscript.

\balance

\bibliographystyle{prsty}

\begin{thebibliography}{10}
	
	\bibitem{2012_JPA_Heiss}
	W.~D. Heiss, Journal of Physics A: Mathematical and Theoretical {\bf 45},
	444016  (2012).
	
	\bibitem{2019_Science_Alu}
	M.-A. Miri and A. Alu, Science {\bf 363},  39  (2019).
	
	\bibitem{2018_Nature_Yoon}
	J.~W. Yoon {\it et~al.}, Nature {\bf 562},  7725  (2018).
	
	\bibitem{2017_Nture_Hodaei}
	H. Hodaei, A. Hassan, S. Wittek, H. Garcia-Gracia, R. El-Ganainy, D.
	Christodoulides, and M. Khajavikhan, Nature {\bf 548},  187  (2017).
	
	\bibitem{2017_Nature_Chen}
	W. Chen, S. Ozdemir, G. Zhao, J. Wiersig, and L. Yang, Nature {\bf 548},  192
	(2017).
	
	\bibitem{2001_PRL_Dembowski}
	C. Dembowski, H. Graf, H. Harney, A. Heine, W. Heiss, H. Rehfeld, and A.
	Richter, Physical Review Letters {\bf 86},  787  (2001).
	
	\bibitem{2003_PRL_Dembowski}
	C. Dembowski, B. Dietz, H. Graf, H. Harney, A. Heine, W. Heiss, and A. Richter,
	Physical Review Letters {\bf 90},  034101  (2003).
	
	\bibitem{2004_PRE_Dembowski}
	C. Dembowski, B. Dietz, H. Graf, H. Harney, A. Heine, W. Heiss, and A. Richter,
	Physical Review E {\bf 69},  056216  (2004).
	
	\bibitem{2016_Nature_Doppler}
	J. Doppler {\it et~al.}, Nature {\bf 537},  76  (2016).
	
	\bibitem{2017_PRA_Zhang}
	X.-L. Zhang, S. Wang, W.-J. Chen, and C. Chan, Physical Review A {\bf 96},
	022112  (2017).
	
	\bibitem{2016_Nature_Xu}
	H. Xu, D. Mason, L. Jiang, and J. Harris, Nature {\bf 537},  80  (2016).
	
	\bibitem{2016_PRX_Ding}
	K. Ding, G. Ma, M. Xiao, Z. Zhang, and C. Chan, Physical Review X {\bf 6},
	021007  (2016).
	
	\bibitem{2011_JPA_Uzdin}
	R. Uzdin, A. Mailybaev, and N. Moiseyev, J. Phys. A: Math. Theor. {\bf 44},
	435302  (2011).
	
	\bibitem{2016_PNAS_Peng}
	B. Peng {\it et~al.}, Proceedings of the National Academy of Sciences {\bf
		113},  6845  (2016).
	
	\bibitem{2019_Optica_Zhen}
	H. Zhou, J.~Y. Lee, S. Liu, and B. Zhen, Optica {\bf 6},  190  (2019).
	
	\bibitem{2019_PRB_Okugawa}
	R. Okugawa and T. Yokoyama, PHYSICAL REVIEW B {\bf 99},  041202(R)  (2019).
	
	\bibitem{2014_PRL_Zhang}
	X. Zhang, C.-L. Zou, L. Jiang, and H.~X. Tang, Physical Review Letters {\bf
		113},  156401  (2014).
	
	\bibitem{2014_PRL_Tabuchi}
	Y. Tabuchi, S. Ishino, T. Ishikawa, R. Yamazaki, K. Usami, and Y. Nakamura,
	Physical Review Letters {\bf 113},  083603  (2014).
	
	\bibitem{2014_PRApplied_Goryachev}
	M. Goryachev, W. Farr, D. Creedon, Y. Fan, M. Kostylev, and M. Tobar, Physical
	Review Applied {\bf 2},  054002  (2014).
	
	\bibitem{2015_PRL_Bai}
	L. Bai, M. Harder, Y. Chen, X. Fan, J. Xiao, and C.-M. Hu, Physical Review
	Letters {\bf 114},  227201  (2015).
	
	\bibitem{2013_PRL_Huebl}
	H. Huebl, C. Zollitsch, J. Lotze, F. Hocke, M. Greifenstein, A. Marx, R. Gross,
	and S. Goennenwein, Physical Review Letters {\bf 111},  127003  (2013).
	
	\bibitem{2017_SciRep_Bhoi}
	B. Bhoi, B. Kim, J. Kim, Y.-J. Cho, and S.-K. Kim, Scientific Reports {\bf 7},
	11930  (2017).
	
	\bibitem{2017_NComm_JQYou}
	D. Zhang, X.-Q. Luo, Y.-P. Wang, T.-F. Li, and J. You, Nature Communications
	{\bf 8},  1368  (2017).
	
	\bibitem{2019_Nature_Lustig}
	E. Lustig, S. Weimann, Y. Plotnik, Y. Lumer, M.~A. Bandres, A. Szameit, and M.
	Segev, Nature {\bf 567},  356  (2019).
	
	\bibitem{2017_PRX_Wang}
	Q. Wang, M. Xiao, H. Liu, S. Zhu, and C. Chan, Physical Review X {\bf 7},
	031032  (2017).
	
	\bibitem{2018_Optica_Yuan}
	L. Yuan, Q. Lin, M. Xiao, and S. Fan, Optica {\bf 5},  1396  (2018).
	
	\bibitem{2018_PRL_Ding}
	K. Ding, G. Ma, Z. Zhang, and C. Chan, Physical Review Letters {\bf 121},
	085702  (2018).
	
	\bibitem{Rogers}
	Rogers Corporation .
	
	\bibitem{SM}
	Supplementary Materials.
	
\end{thebibliography}

\clearpage
\newpage

\begin{center}
	\textbf{Supplementary Material}
\end{center}

\subsection{Device preparation and measurement}
The microwave cavity is prepared on a 1.27-mm-thick TMM10i PCB substrate. The TMM10i substrate has a dielectric constant of 9.8 and is coated with 35-um-thick copper on both sides. The board is cut into the desired size ($12\times5$ mm$^2$) and polished on the four edges using sand paper and cleaned afterward. Silver epoxy is applied to the four edges and cured at $95^\circ$ for 6 hours on a hotplate. The measured resonance frequency is at around 9 GHz as desired. Its resonance linewidth is higher than the air-filled 3D copper cavities because of the high dielectric and metal losses. The TMM10i substrate has a loss tangent of 0.002, and the loss of the silver epoxy is even higher. Besides, since the cavity volume is very small, the SMP probe introduces strong perturbation to the cavity field distribution which also caused additional losses. However, thanks to the significantly reduced cavity volume, strong coupling is still conveniently achievable. The device is measured by a vector network analyzer through out our experiment. Since there is only one SMP probe, reflection measurement is carried out. A circulator is used to separate the reflected signal fro臘୵the input signal to reduce interferences. The reflected signals are collected by Labview programs for post-processing.

\subsection{Parameter extraction and Riemann surface construction}
The Riemann surfaces cannot be directly constructed from the measurement results. They rely on the intrinsic system parameter as list in Eq.(2) of the main text: intrinsic resonance frequency and linewidth for both the magnon and cavity photon modes. These parameters can be numerically extracted from the measured reflection spectra using the following equation:

\begin{equation}\tag{S1}
R(f)=A\times \left\lvert1-\frac{2\kappa_e}{i(f_c-f)+\kappa_c+\frac{g^2}{i(f_m-f)+\kappa_m}}\right\rvert^2,
\label{Eq:reflection}
\end{equation}
\noindent where $R(f)$ is the reflection as a function of frequency $f$, $A$ is the amplitude, $\kappa_e$ is the coupling rate of the cavity with the SMP probe.

\subsection{Anisotropy of the coalesced exceptional point}
As described in the main text, the eigenfrequency has different dependencies on the three parameters: YIG sphere position $y$ and $x$, and the bias magnetic field $H$.  According to Eq.\,(2) in the main text, the square-root term determines the dependence of the eigenfrequency on the system parameters:
\begin{equation}\tag{S2}
\lambda_1=\sqrt{(\delta f-i\Delta\kappa)^2+4g^2}.
\label{Eq:S2}
\end{equation}

\par

\begin{figure}[h]
	\centering
	\includegraphics[width=0.9\linewidth]{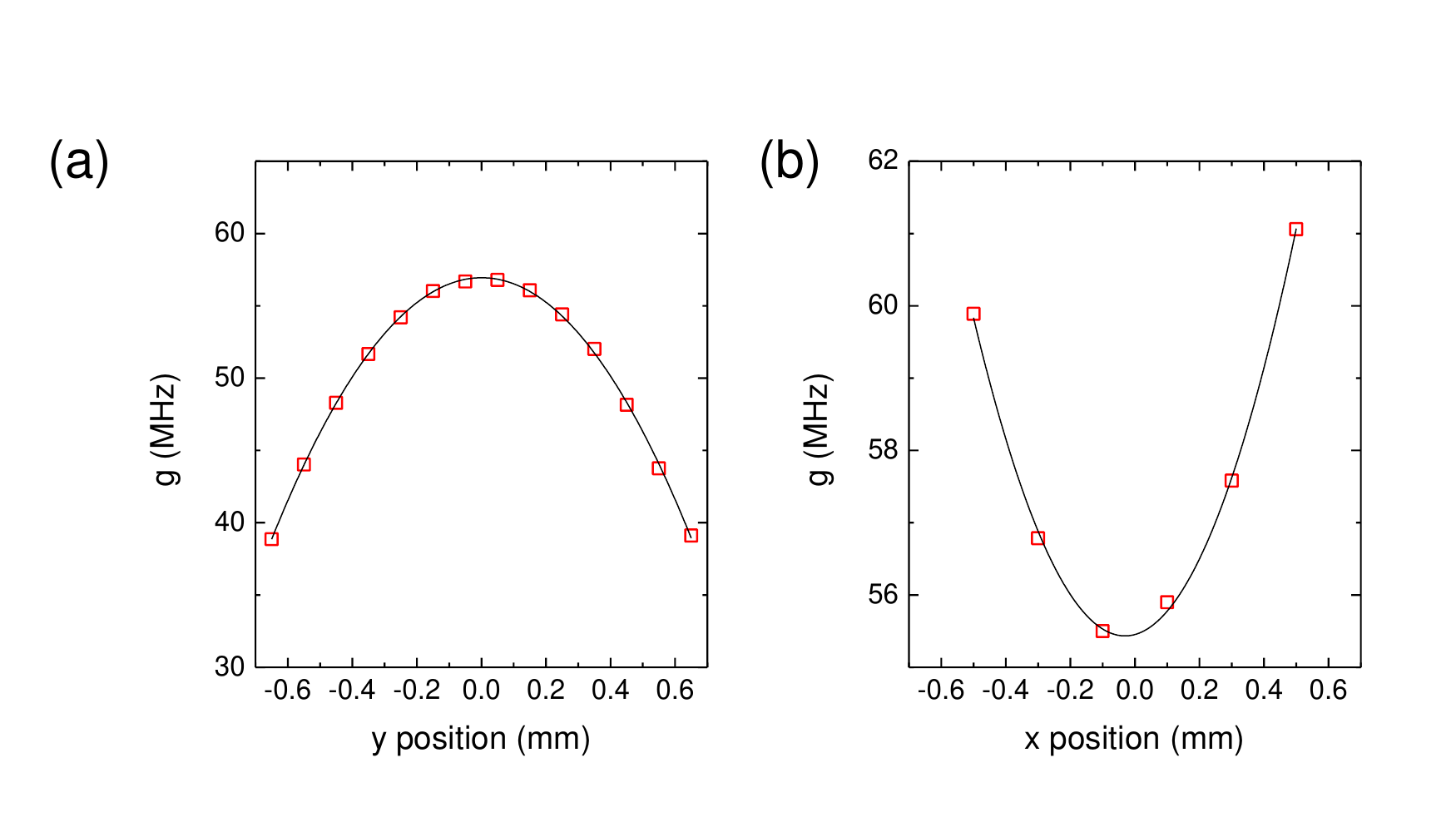}
	\caption{Coupling strength as a function of (a) y position; and (b) x position. Open squares are measurement results. Solid lines are parabolic fittings.}
	\label{fig:figSM2}
\end{figure}

\noindent 1. Dependence on $y$.
In our experiment, the $g-y$ relation is a parabola ($g\propto y^2$), which is confirmed by the numerical fitting (Fig.\,\ref{fig:figSM2}a). The coalescence of EPs occurs at the peak of the parabola.
At the EP, we have $\Delta f=0$, therefore Eq.\,(\ref{Eq:S2}) reduces to $\lambda_1=\sqrt{(-i\Delta\kappa)^2+4g^2}=\sqrt{4g^2-\Delta\kappa^2}$. At the coalesced EP condition, we have $g=g_c=\Delta\kappa/2$. Because this point is the maximum of the $g-y$ parabola, there is always $g\leq \Delta\kappa/2$. So the coupling strength can be expresses as $g=g_c-\delta$, where $\delta$ is a positive number to ensure $g$ does not exceed $g_c$. Under this condition, we have $\lambda_1=\sqrt{4g_c^2+4\delta^2-8g_c\delta-4g_c^2}=2\sqrt{\delta^2-2g_c\delta}$. Under perturbation limit, it can be rewritten as $\lambda_1=2\sqrt{-2g_c\delta}$. Consequently, we have $\mathrm{Im}(\lambda_1)\propto \sqrt{y^2}\propto y$, considering the parabola relation between $g$ and $y$. This explains the linear dependence of the imaginary part of eigenfrequency on $y$ position of the YIG sphere. Since now $\lambda_1$ is purely imaginary, the real part of the eigenfrequency will not be affected and therefore remain constant when sweeping $y$ and is identical (theoretically) for both eigenmodes (Fig.4b in the main text).\\

\noindent 2. Dependence on $x$.
Similarly, the $g-x$ relation in our measurements is also a parabola ($g\propto x^2$ ), as plotted in Fig.\,\ref{fig:figSM2}b. The coalescence of EPs occurs at the bottom of the parabola. At the EP, we have $\Delta f=0$, therefore Eq.\,(\ref{Eq:S2}) reduces to $\lambda_1=\sqrt{(-i\Delta\kappa)^2+4g^2}=\sqrt{4g^2-\Delta\kappa^2}$. At the coalesced EP condition, we have $g=g_c=\Delta\kappa/2$. Because this point is the minimum of the $g-x$ parabola, there is always $g\geq \Delta\kappa/2$. So the coupling strength can be expresses as $g=g_c+\delta$, where $\delta$ is a positive number to ensure $g$ does not go below $g_c$. Under this condition, we have $\lambda_1=\sqrt{4g_c^2+4\delta^2+8g_c\delta-4g_c^2}=2\sqrt{\delta^2+2g_c\delta}$. Under perturbation limit, it can be rewritten as $\lambda_1=2\sqrt{2g_c\delta}$. Consequently, we have $\mathrm{Re}(\lambda_1)\propto \sqrt{x^2}\propto x$, considering the parabola relation between $g$ and $x$. This explains the linear dependence of the real part of eigenfrequency on $x$ position of the YIG sphere. Since now $\lambda_1$ is purely real, the imaginary part of the eigenfrequency will not be affected and therefore remain constant when sweeping $x$ and is identical (theoretically) for both eigenmodes (Fig.4d in the main text).\\

\noindent 3. Dependence on $H$.
To study the dependence on $H$, we take advantage of the fact that $g=g_c=\Delta\kappa/2$ at the EP, and rewrite Eq.\,(\ref{Eq:S2}) as $\lambda_1=\sqrt{\Delta f^2-2i\Delta f\Delta\kappa-\Delta\kappa^2+4g^2}=\sqrt{\Delta f^2-2i\Delta f\Delta\kappa}$. Under perturbation condition, $\Delta f\rightarrow0$, so we have $\lambda_1=\sqrt{-2i\Delta f\Delta\kappa}=(1-i)\sqrt{\Delta f\Delta\kappa}$. Therefore, both the real and imaginary part of $\lambda_1$ have a square-root dependence on $\Delta f$. Because of the linear relation betwen $\Delta f$ and $H$, a square-root dependence on the bias magnetic field $H$ is obtained for both the real and imaginary part of the eigenfrequency.

\end{document}